\documentclass{jpsj3}
\usepackage{txfonts}

\def\Jbf{\mbox{\boldmath $J$}}

\def\Mbf{\mbox{\boldmath $M$}}
\def\Hbf{\mbox{\boldmath $H$}}

\title{Magnetic and Thermal Properties of SmRh$_2$Zn$_{20}$ Single Crystal
}

\author{Yosikazu Isikawa\thanks{E-mail: isikawa@sci.u-toyama.ac.jp}, 
Toshio Mizushima, Aika Fujita, and Tomohiko Kuwai
}

\inst{Graduate School of Science and Engineering, University of Toyama, Toyama 930-8555, Japan }

\abst{
The magnetization, magnetic susceptibility, and specific heat of the single crystalline sample 
SmRh$_2$Zn$_{20}$ were measured.
The valence of Sm ions in SmRh$_2$Zn$_{20}$ was found to be trivalent. 
No evidence of valence fluctuations was detected.
SmRh$_2$Zn$_{20}$ is an antiferromagnet with $T_{\rm N}=$ 2.46 K.
The observed magnetic phase transition temperature in the $C(T,H)$ curves
showed that  $T_{\rm N}$ splits into two in the external field $H$ 
along the [001] and [101] directions.
On the other hand, $T_{\rm N}$ in $H$ along the [111] direction  did not split,
decreasing to 2.20 K at $H=7$ T.
At 2 K, the magnetization $M_{111}$ in $H$ along the [111] direction increased linearly with increasing field,
while $M_{001}$ and $M_{101}$ deviated upward slightly from the linear dependence.
We analyzed the observed magnetic and thermal properties of SmRh$_2$Zn$_{20}$
taking into account the crystalline-electric-field effect, the Zeeman energy, and the exchange interaction. 
The theoretical calculation
well reproduced 
the experimental $\chi(T)$, $M(H)$, $C(T,H)$ and $T_{\rm N}(H)$, suggesting that
the energy scheme of Sm$^{3+}$ is composed of the ground state $\Gamma_7$ and the excited state $\Gamma_8$ with an energy gap of 10.8 K.
The sublattice magnetic moments are expected to be along the $\langle 111\rangle$ direction below $T_{\rm N}$ at $H=0$ T.
Variations of the magnetic structures induced by the external magnetic fields in a narrow temperature region around $T_{\rm N}$ are inferred 
on the basis of  theoretical calculations.
}

\begin{document}
\maketitle

\section{Introduction}  %No sections necessary for express letters, letters and short notes
Many Sm-based compounds have been investigated for more than three decades 
because they show the fundamental physical properties of condensed matters such as valence fluctuation and intermediate valence.\cite{tar1,tar2}
Recently, $RTr_2X_{20}$-type compounds ($R=$ rare earth, $Tr=$ transition metal, $X=$ Al, Zn, Cd)
have attracted much attention because of their various physical properties.\cite{oni1,naga,oni2,tori,isiDy,isiNd}
They crystallize in the cubic structure, and the rare-earth atoms are in the cubic symmetric sites.
 Sakai and Nakatsuji, Higashinaka et al., and Yamada et al. presented 
the 
interesting experimental results\cite{higa,saka,yama} indicating that 
Sm$Tr_2$Al$_{20}$ ($Tr$ = Ti, V, Cr, Ta) are antiferromagnets with strong valence fluctuation,
which brought about 
characteristic behaviors 
such as a large electronic specific-heat coefficient $C/T$, 
a weak temperature dependence of magnetic susceptibility,
a $-$ln $T$-dependent resistivity,
and a field-insensitive phase transition.
The peculiar valence fluctuation is ascribed to the strong $c$-$f$ hybridization.
It is contradictory, however, to the fact that the ground state of Sm ions in all Sm$Tr_2$Al$_{20}$ 
is said to be a quartet $\Gamma_8$ of trivalent Sm. 
Kuwai et al. measured\cite{kuwa} the thermoelectric power $S$ of Sm$Tr_2$Al$_{20}$ 
($Tr$ = Ti, V, Cr)
and found the large values of $\Delta S/T$ at temperatures above and near the N\'eel temperature $T_{\rm N}$.
The large values of $S$ correspond approximately to the large values of $C/T$
since $S$ is proportional to $C$ when both originate from the density of states of the conduction electrons at the Fermi energy.

Sm$Tr_2$Zn$_{20}$ ($Tr$ = Fe, Co, Ru) and Sm$Tr_2$Cd$_{20}$ ($Tr$ = Ni, Pd) have been  investigated by Yazici et al.\cite{yazi} and Jia et al.\cite{jia1}
Sm$Tr_2$Zn$_{20}$ ($Tr$ = Fe, Ru) and SmNi$_2$Cd$_{20}$ exhibit ferromagnetic order, whereas 
SmPd$_2$Cd$_{20}$ is an antiferromagnet and SmCo$_2$Zn$_{20}$ is nonmagnetic down to 110 mK.
The valence of Sm ions in these series is close to trivalent.
The ground state of Sm$^{3+}$ due to the crystalline-electric-field (CEF) effect is quartet $\Gamma_8$ 
for SmRu$_2$Zn$_{20}$\cite{isiSm} and SmPd$_2$Cd$_{20}$\cite{yazi}. 
SmRu$_2$Zn$_{20}$ shows an anomalous magnetic anisotropy of magnetization below the Curie temperature $T_{\rm C}$,
which contradicts the anisotropy predicted from the $\Gamma_8$ ground state.\cite{isiSm}
Isikawa et al. suggested\cite{isiSm} the possibility of the octupole-octupole interaction as a mechanism to explain the anomalous magnetic anisotropy.
Yazici et al. suggested\cite{yazi} that SmRu$_2$Zn$_{20}$ is a rare compound of Sm-based heavy-fermion ferromagnet based on 
the Sommerfeld$-$Wilson and Kadowaki$-$Woods ratios.

A few experimental data revealed 
\cite{taga,take1,take2,take3,take4} that
SmIr$_2$Zn$_{20}$ and SmRh$_2$Zn$_{20}$ are  antiferromagnets with $T_{\rm N}$ at 1.3 and 2.4 K, respectively.
The former has an additional $T_{\rm N}$ at 1.2 K.
The ground state of Sm ions is the $\Gamma_7$ doublet for both compounds,
which are rare examples among the Sm$Tr_2X_{20}$-type compounds.
In SmRh$_2$Zn$_{20}$, the field-induced new phases and the field-induced first-order transition were observed.
However, this first-order transition was sample-sensitive.

In this paper, 
we report the magnetic susceptibility $\chi(T)$, magnetization $M(H)$,
and specific heat $C(T,H)$ of SmRh$_2$Zn$_{20}$ 
to elucidate the fundamental physical properties of the sample.
SmRh$_2$Zn$_{20}$ is an antiferromagnet with a N\'eel temperature $T_{\rm N}$ = 2.46 K.
The observed peak in the $C(T)$ curve at $T_{\rm N}$ is
split into two by the external field $H$, depending on the field direction.
A phase diagram of $T_{\rm N}$ vs $H$ is given.
The experimental results of $\chi(T)$,
$M(H)$, $C(T,H)$, and the field-direction dependence of $T_{\rm N}$ are well reproduced by the
theoretical calculations based on the CEF effect, Zeeman effect, and exchange interaction.
The variations of the magnetic structures in $H$ are discussed on the basis of theoretical calculations.

\section{Experimental Procedure}
Single crystals of SmRh$_2$Zn$_{20}$ and the reference sample YRh$_2$Zn$_{20}$ were grown by the Zn-self-flux method,
which was the same as that described previously.\cite{isiCe,isiSn}
The crystal structure of the cubic CeCr$_2$Al$_{20}$ type was confirmed from the X-ray powder diffraction pattern. 
There was no trace of impurity phases. 
The lattice parameters $a$ of SmRh$_2$Zn$_{20}$ and YRh$_2$Zn$_{20}$ were obtained to be 14.226 and 14.200 \AA, respectively,
which agree with those in the literature\cite{nasc,jiaGd}.
The crystal axis was determined from Laue pictures. 
The samples were shaped using a spark-cutting machine, 
and the weights of the samples SmRh$_2$Zn$_{20}$ and YRh$_2$Zn$_{20}$ are 4.46 and 7.67 mg, respectively,  
which were used for all the measurements. 
We recognized a sample dependence of the physical properties
in SmRh$_2$Zn$_{20}$;
thus, we present here the data of the sample, the specific heat $C(T)$ of which shows the sharpest peak at $T_{\rm N}$ and the highest $T_{\rm N}$. 

The magnetization $M$ and the magnetic susceptibility $\chi$ were
measured at temperatures down to 2.0 K using 
a magnetic property measurement system (MPMS, Quantum Design Inc.).
The specific heat was measured at temperatures down to 0.5 K using a physical property measurement system 
(PPMS, Quantum Design Inc.).

\section{Experimental Results}
\begin{figure}[ht]
\begin{center}
\includegraphics[width=75mm]{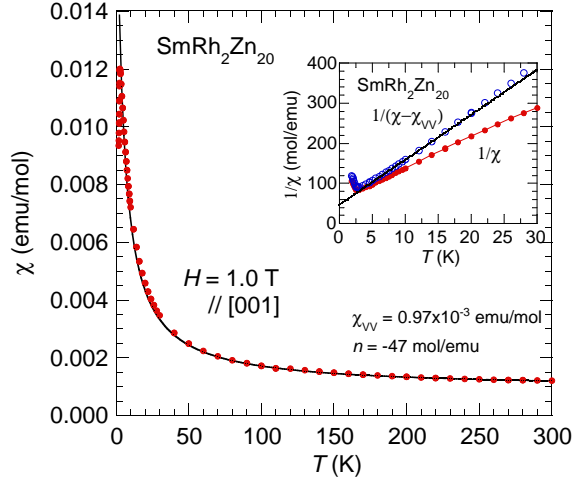}
\caption{\label{SmRh2Zn20_chi100vsT}  (Color online)
Magnetic susceptibility of SmRh$_2$Zn$_{20}$ in the field 1 T along the [001] direction. 
The solid line is a calculated curve. 
Inset: reciprocal magnetic susceptibilities $1/\chi$ (closed red circles) and $1/(\chi-\chi_{\rm VV})$ (blue open circles) in low-temperature region.
The thick solid line in the inset is a calculated curve.
See text for details.}
\end{center}
\end{figure}

Figure \ref{SmRh2Zn20_chi100vsT} shows the temperature dependence of the magnetic susceptibility of SmRh$_2$Zn$_{20}$
in the field 1 T along the [001] direction. 
The value of $\chi$ at 300 K suggests that the Sm ions in SmRh$_2$Zn$_{20}$ are in the trivalent state,
not in the valence-mixing state.\cite{tar1}
The magnetism of the trivalent Sm compound is generally expressed by the sum of two components:
the Curie term that originated from $J=5/2$ and the Van Vleck term that originated from the mixing of $J=5/2$ with 7/2,
that is, $\chi^{3+}(T)= C/T+\chi_{\rm VV}$, where $C$ is the Curie constant of $J=5/2$.
In addition, the magnetic susceptibility is affected by the exchange interaction as follows:
$1/\chi=1/\chi^{3+} -n$.
The solid line in Fig. \ref{SmRh2Zn20_chi100vsT} shows the calculated $\chi(T)$ curve using the two parameters, 
$\chi_{\rm VV}= 0.97\times 10^{-3}$ emu/mol and $n=-47$ mol/emu. 
The experimental susceptibility is in good agreement with the calculated one.
The inset in Fig. \ref{SmRh2Zn20_chi100vsT} shows the reciprocal susceptibility $1/\chi$ 
and corrected reciprocal susceptibility $1/(\chi(T)- \chi_{\rm VV})$
at temperatures below 30 K 
in the same field as in the main figure.
The corrected susceptibility shows a linear dependence of temperature representing the usual Curie$-$Weiss law.
The thick solid line in the inset is the calculated line using $n$.
At low temperatures between 3 and  30 K,
the experimental data is also in good agreement with the calculated one.
The good agreement at temperatures between 3 and 300 K indicates that the CEF effect is small.
The paramagnetic Curie temperature $\theta_{\rm p}$ was deduced to be $-5$ K 
from the data in Fig. \ref{SmRh2Zn20_chi100vsT},
which is related theoretically to the parameter $n$ as $\theta_{\rm p}=nC$.

\begin{figure}[ht]
\begin{center}
\includegraphics[width=75mm]{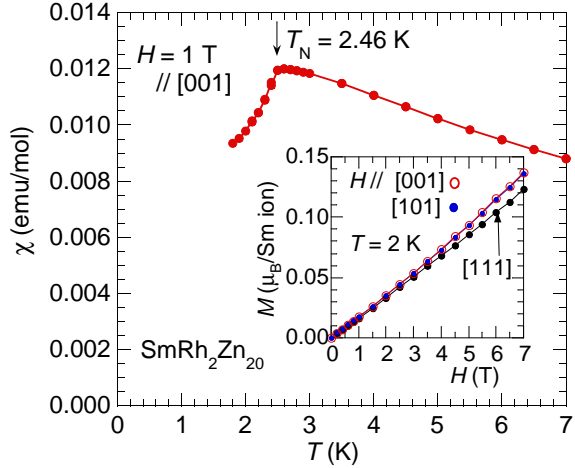}
\caption{\label{SmRh2Zn20_chi100vslowT}  (Color online)
Magnetic susceptibility of SmRh$_2$Zn$_{20}$ in the field 1 T along the [001] direction at temperatures below 7 K.
Inset: magnetization curves at 2 K in the fields along the [001], [101], and [111] directions.}
\end{center}
\end{figure}

Figure \ref{SmRh2Zn20_chi100vslowT} shows the $\chi(T)$ curve
in the field 1 T along the [001] direction at low temperatures.
The magnetic phase transition $T_{\rm N}$ is observed at 2.46 K.
The temperature dependences of $\chi(T)$ along the [101] and [111] directions
were also measured (not shown here)  and they were almost the same as 
that along the [001] direction
at temperatures down to 2 K, implying that 
 $\chi(T)$ does not show any field-direction dependence
even at temperatures below $T_{\rm N}$. 
However, 
the absence of the field-direction dependence of $\chi(T)$ is limited at temperatures in the vicinity of $T_{\rm N}$.
The inset in Fig. \ref{SmRh2Zn20_chi100vslowT} shows the field dependences of magnetization at 2 K
in the fields along the [001], [101], and [111] directions.
The magnetization along the [111] direction shows an almost linear dependence against the field,
but the magnetizations along the [001] and [101] directions gradually deviate upward slightly from the linear dependence at around 2 T.
This result does not indicate that the easy direction of magnetization in fields is parallel to the [001] or [101] direction.

\begin{figure}[ht]
\begin{center}
\includegraphics[width=70mm]{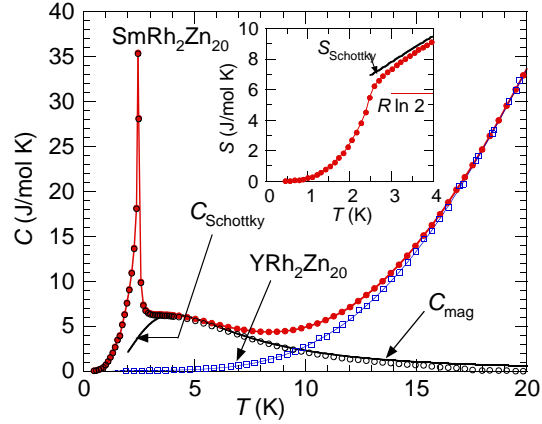}
\caption{\label{SmRh2Zn20_CvsT_20K}  (Color online)
Temperature dependences of the specific heat $C$ of SmRh$_2$Zn$_{20}$ (red closed circles) 
and YRh$_2$Zn$_{20}$ (blue open squares).
$C_{\rm mag}$ denotes the magnetic part of $C$ (black open circles) evaluated as  $C_{\rm SmRh2Zn20}-C_{\rm YRh2Zn20}$.
The thick solid line denoted by $C_{\rm Schottky}$  is  a calculated curve of the Schottky-type specific heat.
Inset: temperature dependence of the entropy of SmRh$_2$Zn$_{20}$. 
The solid line denoted by $S_{\rm Schottky}$ is a calculated curve.
See text for details.  }
\end{center}
\end{figure}

Figure \ref{SmRh2Zn20_CvsT_20K} shows the temperature dependences of the specific heat $C$
of SmRh$_2$Zn$_{20}$ and YRh$_2$Zn$_{20}$.
The sharp peak at $T_{\rm N}$ is observed at 2.46 K, which is the same temperature as that determined in the $\chi(T)$ curve.
The magnetic component $C_{\rm mag}$ of the specific heat is evaluated as $C_{\rm SmRh2Zn20}-C_{\rm YRh2Zn20}$
and is shown by the black open circles in this figure.
The entropy $S$ is numerically calculated by integrating $C/T$ on $T$
and is shown in the inset in Fig. \ref{SmRh2Zn20_CvsT_20K}.
The entropy at $T_{\rm N}$ is approximately equal to $R$ ln 2, 
indicating that the ground state of Sm$^{3+}$ is doublet $\Gamma_7$. 
Moreover, the rapid increase in entropy above $T_{\rm N}$ with increasing temperature suggests that
the excited state $\Gamma_8$ is close to the ground state.
The thick solid line indicated by $C_{\rm Schottky}$ in Fig. \ref{SmRh2Zn20_CvsT_20K}
shows the calculated $C(T)$ curve of SmRh$_2$Zn$_{20}$,
where the energy scheme was assumed to be composed of the doublet ground state and the quartet excited state 
with a gap $\Delta =10.8$ K.
The solid line indicated by  $S_{\rm Schottky}$ in the inset 
is the calculated entropy $S(T)$ curve.
The calculated two lines are in good agreement with the experimental ones above $T_{\rm N}$. 
This narrow energy gap indicates that the CEF effect is weak,
which is consistent with the consideration of the temperature dependence of $\chi(T)$ above $T_{\rm N}$.
$\chi(T)$ was mostly understood on the basis of the Curie$-$Weiss law of Sm$^{3+}$ without the CEF effect. 

\begin{figure}[ht]
\begin{center}
\includegraphics[width=75mm]{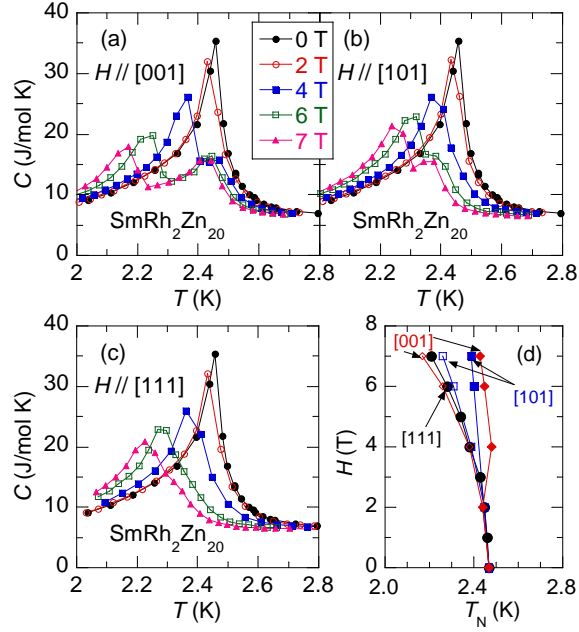}
\caption{\label{SmRh2Zn20_CvsTinHall&TNvsH} (Color online)
Temperature dependences of the specific heat $C$ of SmRh$_2$Zn$_{20}$ in magnetic fields
along the [001] (a),  [101]  (b), and [111] (c) directions.
(d) Magnetic field dependences of $T_{\rm N}$  in the fields 
along the [001]  (filled and open diamonds), [101] (filled and open squares), 
and [111] (filled circles) directions. See text for details.}
\end{center}
\end{figure}

Figures \ref{SmRh2Zn20_CvsTinHall&TNvsH}(a)$-$\ref{SmRh2Zn20_CvsTinHall&TNvsH}(c) 
show the temperature dependences of the specific heat
of SmRh$_2$Zn$_{20}$ in magnetic fields along the [001], [101], and [111] directions.
Interestingly, in the fields along the [001] and [101] directions, 
 $T_{\rm N}$ splits into two and the width of splitting increases with increasing field.
The upper $T_{\rm N}$ does not change with increasing field, whereas the lower one decreases with increasing field.
In the field along the [111] direction,  
 $T_{\rm N}$ gradually decreases without splitting.
Figure \ref{SmRh2Zn20_CvsTinHall&TNvsH}(d) shows the field-direction dependence of  $T_{\rm N}$.
The splitting widths of $T_{\rm N}$ at 7 T are 0.26 K in $H$ // [001] 
and 0.14 K in $H$ // [101].
These of $T_{\rm N}$ changes appear in the narrow temperature range between 2.16 and 2.46 K.

\section{Analysis and Discussion}
\subsection{Hamiltonian}
We discuss the small magnetic anisotropy observed in the magnetization curves at 2 K 
(inset in Fig. \ref{SmRh2Zn20_chi100vslowT}), 
the field-direction dependence of the specific heat $C(T)$, and 
the complex behaviors of $T_{\rm N}$ in fields (Fig. \ref{SmRh2Zn20_CvsTinHall&TNvsH}).
Generally, as a consequence of the characteristics of $\Gamma_7$ and $\Gamma_8$ 
of Sm$^{3+}$ in the cubic symmetry,
the $\Gamma_7$ state is magnetically isotropic, whereas the $\Gamma_8$ state is anisotropic in fields.
Thus, it is expected that this magnetic anisotropy will be caused by the mixing of the $\Gamma_8$ state with the ground state $\Gamma_7$
through the external magnetic field and the exchange field.
We analyze this magnetic anisotropy in the frame of the molecular-field approximation.
For simplicity, a simple antiferromagnetic structure with two sublattices (A and B) is assumed, that is, 
the up- and down-magnetic moments are fixed in the A and B sublattices, respectively.
We define the following Hamiltonian to analyze the magnetic and thermal properties of Sm$^{3+}$ in SmRh$_2$Zn$_{20}$,
\begin{eqnarray}
{\cal H}_i= A_4(O_4^0+5~O_4^4)+  g_J\mu _{\rm B}\Jbf_i\Hbf_{\rm ext}+  g_J\mu _{\rm B}\Jbf_i\Hbf_{\rm mol}(i) ,
\end{eqnarray}
where $i$ denotes the sublattices (A and B). 
The first term in this equation is the CEF energy,\cite{hutc,hut2,llw}
where the six-order term in the CEF energy is excluded
because the quantum number $J$ is  5/2.
The mixing effect with the upper multiplet of $J=7/2$ will be taken into account 
by adding $\chi_{\rm VV}$ to $\chi$ calculated using Eq. (1).
The second term in Eq. (1) is the Zeeman energy due to the external magnetic field $\Hbf_{\rm ext}$,
and the third term is the exchange energy expressed in the molecular-field approximation.
The molecular field $\Hbf_{\rm mol}(i)$ at the $i$ sublattice is expressed using 
the average magnetic moment in the other sublattice $j$
as  $n_{\rm BA}(g_J\mu _{\rm B})\langle\Jbf_j\rangle$, where
$\langle\cdot\cdot\cdot\rangle$ is the thermal average and
$n_{\rm BA}$ is the exchange coupling parameter between the atoms in the A and B sublattices.
The other parameters used in Eq. (1) are conventionally defined.\cite{hutc,llw}

The average magnetization at each sublattice is calculated as
\begin{eqnarray}
\Mbf_i= -g_J\mu _{\rm B}\langle \Jbf_i\rangle
= -g_J\mu _{\rm B}\frac{{\rm Tr}~ \Jbf_i \exp (-\beta {\cal H}_i)}{{\rm Tr}~ \exp (-\beta {\cal H}_i)} ,
\end{eqnarray}
where $\beta=1/k_{\rm B}T$.
The average magnetization $\Mbf$ per atom is $\Mbf=(\Mbf_{\rm A}+\Mbf_{\rm B})/2$.
Note that the physical quantities of $\Jbf_i$, $\Hbf_{\rm mol}$, and $\Hbf_{\rm ext}$ are three-dimensional vectors. 
Thus, Eq. (2) is a set of six equations, which represent the thermal averages of the  $x$-, $y$-, and $z$-components of $\Mbf_{\rm A}$ and $\Mbf_{\rm B}$.
Equation (2) is solved by the iteration method.
By using both values of $\langle\Jbf_{\rm A}\rangle$ and $\langle\Jbf_{\rm B}\rangle$, which have been obtained using Eq. (2),
the specific heat $C$ per mole is evaluated as
\begin{eqnarray}
C= \left(\frac{N_{\rm A}}{2}\right)\frac{\partial }{\partial T}\left(\langle {\cal H}_{\rm A}\rangle +\langle {\cal H}_{\rm B}\rangle
-\frac{1}{2}\langle {\cal H}_{\rm exch}\rangle \right) ,
\end{eqnarray}
where $N_{\rm A}$ is Avogadro's number,
and  the third term in parentheses is used to correct the double counting of the exchange energy 
between the Sm atoms at A and B sublattices.
${\cal H}_{\rm exch}$ in this equation is the same as the third term in Eq. (1)
and is expressed as $n_{\rm BA} (g_J\mu _{\rm B})^2\Jbf_{\rm A} \Jbf_{\rm B}$.

We have only two fitting parameters, i.e., $A_4$ and $n_{\rm BA}$, 
when we numerically calculate $M(T,H)$ and $C(T,H)$ using the above Hamiltonians,
where $H$ is the magnitude of $\Hbf_{\rm ext}$. 
$A_4$ is proportional to $\Delta$ as $A_4= \Delta/360$ in the unit [K],
and $n_{\rm BA}$ is physically the same as $n$ [mol/emu] through the relationship of 
$n_{\rm BA} =n N_{\rm A}$. 
As already mentioned in Sect. 3, 
the energy scheme of Sm$^{3+}$ is composed of the doublet ground state $\Gamma_7$ and the quartet excited state $\Gamma_8$
with $\Delta=$ 10.8 K. Thus, $A_4$ is set to be 0.030 [K].
The value of $n_{\rm BA}$ was accurately determined to be $-1.45~k_{\rm B}/(g_J\mu_{\rm B})^2$
by adjusting the calculated $T_{\rm N}$ to the experimental $T_{\rm N}$.
Consequently, the sublattice magnetic moments $M_{\rm A}$ and $M_{\rm B}$ are evaluated 
as a result of the iteration process of Eq. (2)
using the two parameters $A_4$ and $n_{\rm BA}$.
This means that the magnetic alignments of each sublattice moment in $H$ are inferred.
First, we present the calculated results of $\chi(T)$, $M(H)$, $C(T,H)$, and $T_{\rm N}(H)$
in comparison with the respective experimental data in the next subsection.
After that, we discuss the magnetic alignments of sublattice magnetic moments in $H$.

We have to note one important result concluded from this CEF calculation.
At temperatures below $T_{\rm N}$,
the magnetic moments at the sublattices are along the $\langle 111\rangle$ direction at $H=0$,
namely, there exist four magnetic domains below $T_{\rm N}$.    
For simplicity, we have calculated 
the $\chi(T)$, $M(H)$, and  $C(T)$ curves in fields for a single domain,
in which the up- and down-magnetic moments are along the [111] and $[\bar{1}~\bar{1}~\bar{1}]$  directions, respectively.

\subsection{Calculation and comparison with the experimental data}
\begin{figure}[ht]
\begin{center}
\includegraphics[width=70mm]{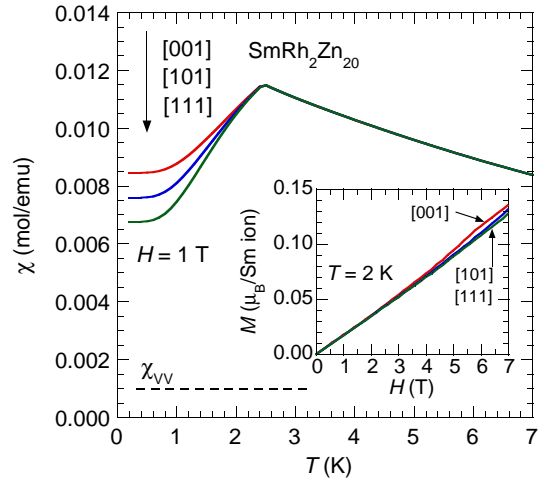}
\caption{\label{SmRh2Zn20_cal_chivslowT_MvsH} (Color online)
Temperature dependences of the calculated magnetic susceptibility of SmRh$_2$Zn$_{20}$ in the field 1 T along the [001], [101], and [111] directions.
Inset: calculated magnetization curves at 2 K along the three directions.
See text for details.}
\end{center}
\end{figure}

Figure \ref{SmRh2Zn20_cal_chivslowT_MvsH} shows the calculated temperature dependences of 
the magnetic susceptibility $\chi$ of
SmRh$_2$Zn$_{20}$ in the field 1 T along the [001], [101], and [111] directions,
where $\chi$ has been shifted by the amount of $\chi_{\rm VV}$
because the mixing effect with the multiplet of $J=7/2$ is not taken into account in Eq. (2). 
In the temperature range between 2 K and $T_{\rm N}$, the magnetic anisotropy is very small, which is consistent with the experimental results.
Below 2 K, however, the calculated $\chi$ shows a magnetic anisotropy, 
although we cannot compare it with the experimental data, which is limited above 2 K.

The inset in Fig. \ref{SmRh2Zn20_cal_chivslowT_MvsH} shows the calculated magnetization curves at 2 K in the fields along the [001], [101], and [111] directions.
$M_{111}(H)$ in $H$ // [111] increases linearly with increasing field without any anomalies such as spin flops,
which is in good agreement with the experimental behavior.
On the other hand, the $M_{101}(H)$ curve deviates slightly from the $M_{111}(H)$ curve around 3 T,
and the $M_{001}(H)$ curve deviates further from the $M_{111}(H)$ curve.
The characteristic behaviors of the three calculated $M(H)$ curves at 2 K 
are in good agreement with the experimental ones.

\begin{figure}[ht]
\begin{center}
\includegraphics[width=75mm]{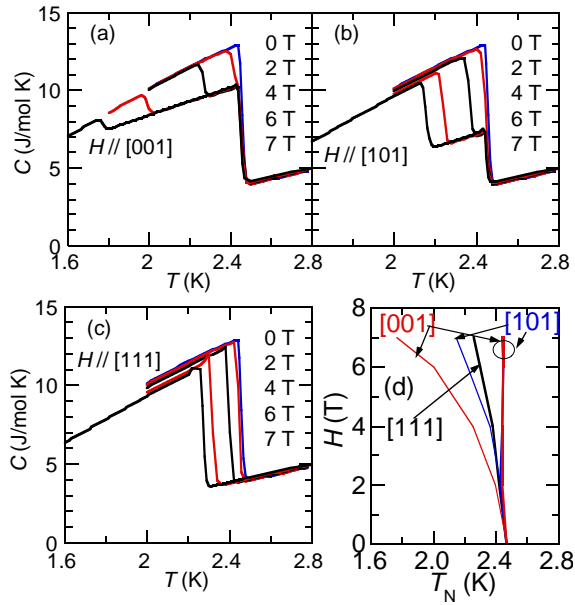}
\caption{\label{SmRh2Zn20_cal_CvsT_Hall&TNvsH} (Color online)
Temperature dependences of the calculated specific heat of SmRh$_2$Zn$_{20}$ in the fields along the [001]  (a),
 [101] (b), and  [111] (c) directions.
(d) Calculated field dependences of $T_{\rm N}$. See text for details.}
\end{center}
\end{figure}

Figures \ref{SmRh2Zn20_cal_CvsT_Hall&TNvsH}(a)$-$\ref{SmRh2Zn20_cal_CvsT_Hall&TNvsH}(c) show the calculated specific heat $C$ in the fields of 
0, 2, 4, 6, and 7 T
along the [001], [101], and [111] directions.
In the case of $H$ // [001] and [101],  $T_{\rm N}$ splits into two, 
and the width of the splitting increases with increasing field.
In $H$ // [111],  $T_{\rm N}$ does not split but moderately decreases with increasing $H$.
In Fig. \ref{SmRh2Zn20_cal_CvsT_Hall&TNvsH}(d), the external-field dependences of $T_{\rm N}$ are plotted.
The splitting widths of $T_{\rm N}$ at 7 T are 0.7 K in $H$ // [001], 0.3 K in $H$ // [101], and zero in $H$ // [111].
These characteristic behaviors are also in good agreement with the experimental ones.
However, note that the height of the calculated specific heat at $T_{\rm N}$ at 0 T is one-third 
compared with that of the experimental specific heat at $T_{\rm N}$,
and the second-peak height separated by $H$ is also lowered compared with the experimental one.
This is caused by the fact that the numerical calculation is based on the molecular-field approximation.
The magnetic fluctuation 
$\langle (\Jbf_{\rm A}-\langle \Jbf_{\rm A}\rangle)(\Jbf_{\rm B}-\langle\Jbf_{\rm B}\rangle)\rangle$, 
which is more important in the vicinity of $T_{\rm N}$ than at temperatures far from $T_{\rm N}$,
cannot be taken into account in this approximation.
In fact, at 2.0 and 2.8 K, 
which are far below and far above $T_{\rm N}$, respectively,
the values of the calculated $C$ are approximately equal to those of the experimental $C$.

It is concluded that
the theoretical calculations have successfully reproduced the macroscopic experimental data, 
$\chi(T)$, $M(T,H)$, $C(T,H)$, and $T_{\rm N}(H)$, by using only two parameters, $A_4$ and $n_{\rm BA}$.

\subsection{Alignment of sublattice magnetic moments}
In this subsection, we present 
the microscopic magnetic structures of SmRh$_2$Zn$_{20}$ in the vicinity of $T_{\rm N}$, which have been
deduced from the model calculations.
Here, we discuss the single magnetic domain.

\begin{figure}[ht]
\begin{center}
\includegraphics[width=80mm]{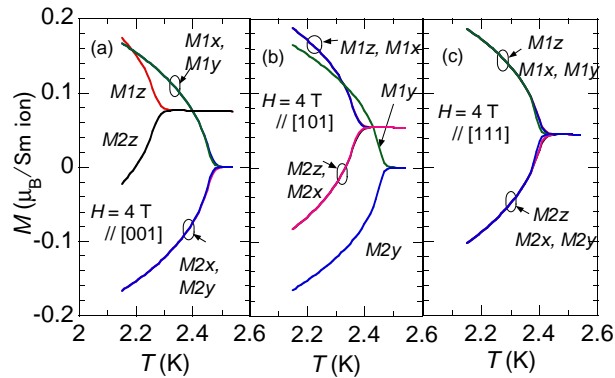}
\caption{\label{SmRh2Zn20_cal_MzxyvsT} (Color online)
Temperature dependences of the calculated $x$-, $y$-, and $z$-components of respective sublattice magnetic moments of SmRh$_2$Zn$_{20}$ in the field 4 T along the [001] (a),  [101] (b), and [111]  (c) directions.
The subscripts 1 and 2 denote the up- and down-sublattice moments, respectively.}
\end{center}
\end{figure}

Figures \ref{SmRh2Zn20_cal_MzxyvsT}(a)$-$\ref{SmRh2Zn20_cal_MzxyvsT}(c) show the temperature dependences of the calculated 
$x$-, $y$-, and $z$-components of the respective sublattice magnetic moments of SmRh$_2$Zn$_{20}$ in the field 4 T.
The subscripts 1 and 2 denote the up- and down-sublattice moments, respectively.
In $H$ along the [001] direction, as seen in Fig. \ref{SmRh2Zn20_cal_MzxyvsT}(a),
the lower $T_{\rm N}$ is found to be the temperature at which the antiferromagnetic alignment
along the [001] direction ($z$-component) disappears,
and the upper $T_{\rm N}$ is  the temperature at which the antiferromagnetic alignment 
of both the $x$- and $y$-components disappears.
In $H$ along the [101] direction [Fig. \ref{SmRh2Zn20_cal_MzxyvsT}(b)],
the lower $T_{\rm N}$ is the temperature at which the antiferromagnetic alignment along the [101] direction disappears,
and the upper $T_{\rm N}$ is the temperature at which the antiferromagnetic alignment 
of the $y$-component disappears.
In $H$ along the [111] direction [Fig. \ref{SmRh2Zn20_cal_MzxyvsT}(c)], 
the magnetizations $\Mbf_{\rm A}$ and $\Mbf_{\rm B}$ are parallel to the [111] direction, and
the antiferromagnetic alignment disappears at $T_{\rm N}$. Thus,
the magnetic susceptibility along the [111] direction can be considered  a parallel magnetic susceptibility.
In the cases of $H$ along the [001] and [101] directions,
the antiferromagnetic component along the $H$ directions disappears 
at temperatures between the lower and upper $T_{\rm N}$'s.
Thus, the magnetic susceptibility along these directions can be considered a perpendicular magnetic susceptibility.
The lower and upper $T_{\rm N}$'s as well as the  $T_{\rm N}$ along the [111] direction are  
 second-order phase-transition temperatures.

\begin{figure}[ht]
\begin{center}
\includegraphics[width=80mm]{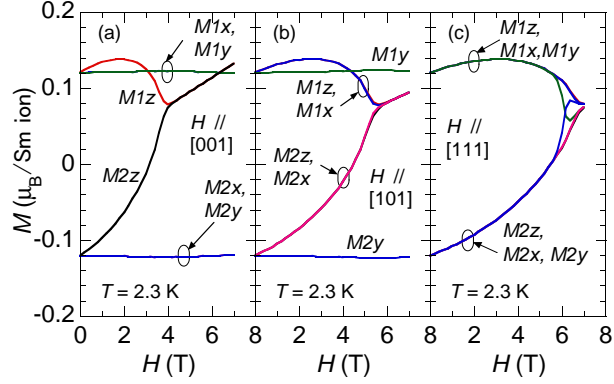}
\caption{\label{SmRh2Zn20_cal_MzxyvsH} (Color online)
Field dependences of the calculated $x$-, $y$-, and $z$-components of respective sublattice magnetic moments of SmRh$_2$Zn$_{20}$ at
2.3 K in fields along the [001]  (a),  [101]   (b), and [111] (c) directions.
The subscripts 1 and 2 denote the up- and down-sublattice moments, respectively.}
\end{center}
\end{figure}

Figures \ref{SmRh2Zn20_cal_MzxyvsH}(a)$-$\ref{SmRh2Zn20_cal_MzxyvsH}(c) show the field dependences of the calculated 
$x$-, $y$-, and $z$-components of the respective sublattice magnetic moments of SmRh$_2$Zn$_{20}$ at 2.3 K.
In the cases of $H$ along the [001] and [101] directions, 
as seen in Figs. \ref{SmRh2Zn20_cal_MzxyvsH}(a) and \ref{SmRh2Zn20_cal_MzxyvsH}(b), 
the antiferromagnetic alignments along the $H$ directions disappear near 4 and 5 T, respectively.
At these points, any anomalous discontinuity such as spin flopping is not theoretically concluded.
On the other hand, the antiferromagnetic alignments perpendicular to the $H$ directions are robust against $H$ as seen in the figures.
In  $H$ along the [111] direction [Fig. \ref{SmRh2Zn20_cal_MzxyvsH}(c)], 
the magnetic moments $\Mbf_{\rm A}$ and $\Mbf_{\rm B}$ are  parallel to the $H$ direction 
below 6 T. Thus, the magnetization along this direction is due to the parallel susceptibility,
as mentioned previously.

The susceptibility $\chi_{111}$ in $H$ along the [111] direction has been considered the parallel susceptibility $\chi_{//}$.
However, as seen in Fig. \ref{SmRh2Zn20_cal_chivslowT_MvsH},
the calculated $\chi_{111}(T)$ does not approach zero at 0 K. 
This behavior is contrary to the common behaviors of usual antiferromagnets.
We briefly comment on the discrepancy between the calculated $\chi_{//}(T)$ and the $\chi_{//}(T)$ for usual antiferromagnets,  
although the experimental $\chi(T)$ is limited above 2 K. 
In the case of usual antiferromagnets, the sublattice moments $\Mbf_{\rm A}$ and $\Mbf_{\rm B}$ are saturated at 0 K,
and they do not change  even if $H$ is applied;
thus, $\chi_{//}$ is equal to zero.
However, when $\Delta \neq 0$, 
the sublattice moments $\Mbf_{\rm A}$ and $\Mbf_{\rm B}$ are not saturated at 0 K,
that is, the moment $\Mbf_{\rm A}$ increases and $\Mbf_{\rm B}$  decreases with the  application of  the external field 
via a Van Vleck-type mechanism between $\Gamma_7$ and $\Gamma_8$ states in Sm$^{3+}$;
thus, $\chi_{//}$ is not equal to zero.

The microscopic magnetic structures deduced from the theoretical calculations
have provided an insight into  SmRh$_2$Zn$_{20}$.
To confirm the magnetic structures of SmRh$_2$Zn$_{20}$, 
neutron scattering experiments should be performed.

\subsection{Conclusions}
We prepared single crystals of SmRh$_2$Zn$_{20}$ and measured  $\chi(T)$, $M(T,H)$, and $C(T,H)$.
Sm ions in SmRh$_2$Zn$_{20}$ are in the trivalent state, not in the valence fluctuating state.
SmRh$_2$Zn$_{20}$ is an antiferromagnet with $T_{\rm N}=$ 2.46 K.
$T_{\rm N}$ splits into two in $H$ // [001] and [101].
The upper $T_{\rm N}$ does not change in fields up to the maximum $H$ of 7 T, 
whereas the lower $T_{\rm N}$ decreases. 
In $H$ // [111],  $T_{\rm N}$ does not split
and decreases to 2.20 K at 7 T.
At 2 K, the magnetization $M_{111}$ along the [111] direction increases linearly with increasing field,
whereas $M_{001}$ and $M_{101}$ deviate upward slightly from the linear dependence of $M_{111}$.
We analyzed these magnetic and thermal properties of SmRh$_2$Zn$_{20}$
taking into account the CEF effect,  Zeeman energy, and  exchange interaction. 
The theoretical calculations have successfully reproduced
the experimental $\chi(T)$, $M(H)$, $C(T,H)$, and $T_{\rm N}(H)$. 
The energy scheme of Sm$^{3+}$ obtained is composed of the ground state $\Gamma_7$ and the excited state $\Gamma_8$ with the energy gap of 10.8 K.
The sublattice magnetic moments are along the $\langle 111\rangle$ direction below $T_{\rm N}$ at $H=0$ T.
The magnetic structures in magnetic fields in the temperature region between the split $T_{\rm N}$'s could be inferred on the basis of the theoretical calculations.

\section*{Acknowledgment}
We would like to thank 
K. Nishimura 
for the critical reading of our manuscript and 
many useful discussions.

 %\section*{References}

\vfill

\begin{thebibliography}{9}%
\bibitem{tar1}  J. M. Tarascon, Y. Isikawa, B. Chevalier, J. Etourneau, P. Hagenmuller, and M. Kasaya, J. Phys. (Paris)
{\bf 41}, 1135 (1980).
\bibitem{tar2} J. M. Tarascon, Y. Isikawa, B. Chevalier, J. Etourneau, P. Hagenmuller, and M. Kasaya, J. Phys. (Paris)
{\bf 41}, 1141 (1980).
\bibitem{oni1}  T. Onimaru, K. T. Matsumoto, Y. F. Inoue, K. Umeo, Y. Saiga, Y. Matsushita, R. Tamura, K. Nishimoto, I. Ishii, T. Suzuki, and T. Takabatake, J. Phys. Soc. Jpn. {\bf 79}, 033704 (2010).
\bibitem{naga}  N. Nagasawa, T. Onimaru, K. T. Matsumoto, K. Umeo, and T. Takabatake, J. Phys.: Conf. Ser. {\bf 391}, 012051 (2012).
\bibitem{oni2}  T. Onimaru, N. Nagasawa, K. T. Matsumoto, K. Wakiya, K. Umeo, S. Kittaka, T. Sakakibara, Y. Matsushita, and T. Takabatake, Phys. Rev. B {\bf 86},   184426 (2012). 
\bibitem{tori}  M. S. Torikachvili, S. Jia, E. D. Mun, S. T. Hannahs, R. C. Black, W. K. Neils, D. Martien, S. L. Bud'ko, and P. C. Canfield, Proc. Natl. Acad. Sci. U.S.A. {\bf 104}, 9960 (2007).
\bibitem{isiDy}	Y. Isikawa, T. Mizushima, S. Miyamoto, K. Kumagai, M. Nakahara, H. Okuyama, T. Tayama, and T. Kuwai,  J. Korean Phys. Soc. {\bf 63}, 644 (2013).
\bibitem{isiNd}  Y. Isikawa, J. Ejiri, T. Mizushima, and T. Kuwai, J. Phys. Soc. Jpn.  {\bf 82}, 123708 (2013). 
\bibitem{saka}  A. Sakai and S. Nakatsuji, Phys. Rev. B {\bf 84}, 201106(R) (2011).
\bibitem{higa}  R. Higashinaka, T. Maruyama, A. Nakama, R. Miyazawa, Y. Aoki, and H. Sato, J. Phys. Soc. Jpn.  {\bf 80}, 093703 (2011). 
\bibitem{yama}  A. Yamada, R. Higashinaka, R. Miyazaki, K. Fushiya, T. D. Matsuda, Y. Aoki, W. Fujita, H. Harima, and H. Sato, J. Phys. Soc. Jpn. {\bf 82}, 123710 (2013). 
\bibitem{kuwa}  T. Kuwai, T. Furuyama, K. Tada, T. Mizushima, and Y. Isikawa, J. Phys. Soc. Conf. Proc. {\bf 3}, 011040 (2014).
\bibitem{yazi}  D. Yazici, B. D. White, P.-C. Ho, N. Kanchanavatee, K. Huang, A. J. Friedman, A. S. Wong, V. W. Burnett, N. R. Dilley, and M. B. Maple, Phys. Rev. B {\bf 90}, 144406 (2014).
\bibitem{jia1} S. Jia, N. Ni, S. L. Bud'ko, and P. C. Canfield, Phys. Rev. B {\bf 80}, 104403 (2009).
\bibitem{isiSm}  Y. Isikawa, T. Mizushima, J. Ejiri, and T. Kuwai, J. Phys. Soc. Jpn.  {\bf 83}, 073701 (2014). 
\bibitem{taga}  Y. Taga, K. Sugiyama, K. Enoki, Y. Hirose, K. Iwakawa, A. Mori, K. Ishida, T. Takeuchi, M. Hagiwara, K. Kindo, R. Settai, and Y. Onuku, J. Phys. Soc. Jpn.  {\bf 81}, SB051 (2012).
\bibitem{take1}  M. Tanahashi, K. Adachi, T. Sasahara, N. Kase, T. Nakano, and N. Takeda, 
Abstr. Meet. Physical Society of Japan (2015 Autumn Meet.), Part 3, p. 1858, 17aPS64 [in Japanese].
\bibitem{take2}  M. Tanahashi, K. Adachi, T. Sasahara, N. Kase, T. Nakano, N. Takeda, Y. Kono, and T. Sakakibara, 
Abstr. Meet. Physical Society of Japan (2015 Annual Meet.), Part 3, p. 2184, 22aPS94 [in Japanese].
\bibitem{take3}   K. Adachi, T. Sasahara, N. Kase, T. Nakano, and N. Takeda, 
Abstr. Meet. Physical Society of Japan (2014 Annual Meet.), Part 3, p. 623, 28aPS132 [in Japanese].
\bibitem{take4}  T. Sasahara,  T. Nakano, and N. Takeda, 
Abstr. Meet. Physical Society of Japan (2013 Annual Meet.), Part 3, p. 626, 27aPS69 [in Japanese].
\bibitem{isiCe}  Y. Isikawa, T. Mizushima, K. Kumagai, and T. Kuwai, J. Phys. Soc. Jpn.  {\bf 82}, 083711 (2013). 
\bibitem{isiSn}  Y. Isikawa, T. Mizushima,  J. Ejiri, S. Kitayama, K. Kumagai, T. Kuwai, P. Bordet, and P. Lejay, J. Phys. Soc. Jpn.  {\bf 84}, 074707 (2015). 
\bibitem{nasc}  T. Nasch, W. Jeitschko, and U. C. Rodewald, Z. Naturforsch. B {\bf 52}, 1023 (1997).
\bibitem{jiaGd} S. Jia, N. Ni, G. D. Samolyuk, A. Kracher, K. Dennis, H. Ko, G. J. Miller, S. L. Bud'ko, and P. C. Canfield, Phys. Rev. B {\bf 77}, 104408 (2008).
\bibitem{hutc}  M. T. Hutchings, Solid State Phys. {\bf 16}, 227 (1964).
\bibitem{llw}   K. R. Lea, M. J. M. Leask, and W. P. Wolf, J. Phys. Chem. Solids, {\bf 23}, 1381 (1962). 
\bibitem{hut2}  In the notation of Lea et al.,\cite{llw} $A_4$ is equal to $Wx/60$.

\vfill
\end{thebibliography}
\end{document}